\documentclass[journal,10pt]{IEEEtran}
\usepackage{amsmath,amsfonts}

\usepackage{algorithmic}
\usepackage{algorithm}
\usepackage{array}
\usepackage{amsmath}
\usepackage{textcomp}
\usepackage{stfloats}
\usepackage{url}
\usepackage{verbatim}
\usepackage{graphicx}
\usepackage{cite}
\usepackage{xcolor}
\usepackage{subfigure}
\usepackage{booktabs}
\usepackage{multirow}
\usepackage{enumitem}  % 自定义列表
\usepackage{cuted} % 导言区添加
\usepackage{array}
\usepackage{amssymb}
\usepackage{bigstrut}
\usepackage{makecell}
\usepackage{adjustbox}
\usepackage{booktabs}
\usepackage{pdfpages}  % 用于插入 PDF 页面
\begin{document}

\title{A Novel Site-Specific Inference Model for Urban Canyon Channels: From Measurements to Modeling}

\author{Junzhe Song,~\IEEEmembership{Student Member,~IEEE,}
Ruisi He,~\IEEEmembership{Senior Member,~IEEE,}
Mi Yang,~\IEEEmembership{Member,~IEEE,}\\
Zhengyu Zhang,~\IEEEmembership{Student Member,~IEEE,}
Xinwen Chen,
Xiaoying Zhang,
Bo Ai,~\IEEEmembership{Fellow,~IEEE,}

\thanks{
J. Song, R. He, M. Yang, Z. Zhang, X. Chen, and B. Ai are with the Schooof Electronics and Information Engineering, Beijing Jiaotong University, Beijing 100044, China (email: 25115063@bjtu.edu.cn; ruisi.he@bjtu.edu.cn; myang@bjtu.edu.cn; 21111040@bjtu.edu.cn; 15162828600@163.com; boai@bitu.edu.cn)

X. Zhang is with the College of Electronic Science and Technology, National University of Defense Technology, Changsha 410073, China (email: zhangxiaoying@nudt.edu.cn).

% Bingcheng Liu is with Aerospace Information Research Institute, Chinese Academy of Sciences, Beijing 100094, China (e-mail: liubc@aircas.ac.cn).

% Jiahui Han and Haoxiang Zhang are with China Academy of Industrial Internet, Ministry of Industry and Information Technology, Beijing, China (email: hjh1760708@126.com; zhx61778294@126.com).

}}

\maketitle

\begin{abstract}
With the rapid development of intelligent transportation and smart city applications, urban canyon has become a critical scenario for the design and evaluation of wireless communication systems. Due to its unique environmental layout, the channel characteristics in urban canyon are strongly a street geometry and building distribution, thereby exhibiting significant site-specific channel condition. However, this feature has not been well captured in existing channel models. In this paper, we propose a site-specific channel inference model based on environmental geometry, the model is parameterized using sub-6GHz channel measurements. Multipath components (MPCs) are extracted and clustered according to geometric propagation, which are explicitly derived from the influence of canyon width, thereby establishing an interpretable mapping between the physical environment and statistical characteristics of MPCs. A step-by-step implementation scheme is presented. Subsequently, the proposed site-specific channel inference model is validated by comparing second-order statistics of channels, derived from the model and measurements. The results show that the proposed model achieves high accuracy and robustness in different urban canyon scenarios.
\end{abstract}


\IEEEdisplaynontitleabstractindextext

\IEEEpeerreviewmaketitle

\section{Introduction}

\IEEEPARstart{U}rban canyon environments have emerged as one of the critical scenarios in the design and deployment of next-generation wireless communication systems. With the advancement of 5G and the evolution towards 6G \cite{huang2022artificial}, the demands of applications such as intelligent transportation, vehicle-to-everything (V2X) \cite{he2019propagation}, and smart cities are continuously increasing, making reliable communication in dense urban areas a fundamental requirement \cite{vardhan2025aber,tian2025analytical}. Characterized by tall buildings and narrow streets, urban canyons present highly complex propagation environments where signals are severely affected by reflections, diffractions, and blockages \cite{he2015characterization,zhang2025non}.

In general, urban canyon channels exhibit the following characteristics \cite{lyu2019characterizing}: i) \textit{pronounced multipath effects:} Reflectors such as buildings and walls generate a large number of discrete multipath components (MPCs), leading to significantly greater delay spread and angular spread compared to open areas. ii) \textit{strong time variability and non-stationarity:} The movement of the transmitter and receiver, together with the presence of dynamic scatterers in the environment, causes the channel characteristics to evolve rapidly over time. iii) \textit{non-uniform angular spreads:} In urban canyon scenarios, the azimuthal spread is usually much larger than the elevation spread, while intersections introduce additional scattering paths. iv) \textit{remarkable environmental dependence:} Channel characteristics are highly dependent on street geometry and building distribution, exhibiting strong site-specific correlations. However, existing standardized models, such as the 3GPP channel models \cite{zhu20213gpp}, WINNER II \cite{kyosti2007winner}, COST \cite{liu2012cost,he2025cost}, and the map-based METIS model\cite{unknown}, typically inherit conventional cellular channel modeling frameworks, making only limited parameter adjustments (e.g., reducing base station height) \cite{guan2021towards,peter2016channel}. Therefore, channel models capable of accurately capturing urban environmental features and faithfully reflecting the propagation mechanisms are essential.

Site-specific channel models represent a class of modeling approaches that incorporate the geometrical features of the environment into the channel characterization process \cite{10742569}. These models can be constructed through ray-tracing simulations, measurement-driven approaches \cite{10872967}, or hybrid frameworks that combine both. In urban canyon scenarios, where signal propagation is strongly governed by environment geometry, site-specific modeling becomes essential. However, most existing site-specific channel models focus on simulating the propagation environment, but they rarely establish a direct inference relationship between environmental parameters and channel characteristics. They typically treat the environment as explicit input to deterministic simulations, rather than learning the underlying mapping that links geometry to channel statistics. Such a limitation hinders the generalization capability of current models, as new environments often require extensive measurements or time-consuming simulations. To overcome this challenge, it is essential to develop a site-specific inference model that leverages environmental geometry parameters as the input and predicts corresponding channel characteristics. By doing so, one can extend channel generation to arbitrary urban canyon scenarios without relying on exhaustive measurements or full-scale ray-tracing, thereby providing a more scalable and flexible modeling framework for next-generation wireless communication systems \cite{he2024wireless}.
% During recent decades, both academia and industry have proposed a variety of standardized channel models that incorporate urban scenarios, such as the 3GPP channel models \cite{zhu20213gpp}, WINNER II \cite{kyosti2007winner}, COST 2100 \cite{liu2012cost}, and the map-based METIS model\cite{unknown}. However, despite their solid foundations, existing standardized models remain insufficient in capturing the unique propagation characteristics of urban canyon environments \cite{guan2021towards,peter2016channel}. This motivates the development of site-specific channel models that integrate environmental features into the modeling process.

Several measurements and models for urban canyon channels have been carried out \cite{kanhere2024calibration,10742569,rainer2021scalable,chizhik2023accurate,adhikari2025around,huang2020non,li2022non,guo2024characterization,yang2023dynamic,hammoud2024double,bignotte2022measurement,yusuf2021autoregressive,bock2025physics,bian2021general,radpour2023dynamic,assiimwe2022mobility,huang2020geometry,huang2022geometry,huang2023mixed}. These models mainly fall into three categories: i) \textit{geometry-based deterministic channel models (GDCMs):} These models characterize MPC parameters in a fully deterministic manner, typically relying on ray-tracing methods \cite{kanhere2024calibration}. Their accuracy depends on detailed knowledge of building layouts, street geometries, and material properties \cite{10742569}. However, the strong reliance on environmental information limits their applicability in large-scale simulations, real-time evaluations, or scenarios with incomplete environmental data \cite{rainer2021scalable}. ii) \textit{nongeometrical stochastic channel models}: These models determine the physical parameters of canyon channels without considering any underlying geometrical structures. For instance, \cite{chizhik2023accurate,adhikari2025around,huang2020non,li2022non,guo2024characterization,yang2023dynamic,hammoud2024double,bignotte2022measurement,yusuf2021autoregressive} introduce general stochastic models. Such models typically generate channel characteristics (e.g., delay, angle, and power) directly based on statistical distributions, but they lack explicit correspondence to the physical environment, making it difficult to interpret or capture scenario-specific variations \cite{bock2025physics}. iii) \textit{geometry-based stochastic channel models (GSCMs)}: These models describe the random distribution of effective scatterers and apply simplified ray tracing to obtain the channel impulse response. Depending on the scatterer deployment strategy, GSCMs can be further classified into two types: regular-shaped GSCMs (RS-GSCMs), e.g., \cite{bian2021general,yuan2013novel,radpour2023dynamic,assiimwe2022mobility}, and irregular-shaped GSCMs (IS-GSCMs), e.g., \cite{huang2020geometry,huang2022geometry,huang2023mixed}. Nevertheless, existing GSCMs often assume scatterers to be uniformly or specifically distributed within canonical geometrical regions (e.g., circular rings, ellipses, or strips) rather than directly incorporating real environmental information, thereby limiting their fidelity in realistic environments.

Moreover, several innovative studies on site-specific urban canyon channels have overcome the limitations of traditional modeling approaches. For example, Gupta et al. conducted large-scale 28 GHz field measurements in the Manhattan street canyons of New York and trained a data-driven path loss predictor by integrating LiDAR-based street clutter and building grid features \cite{gupta2022machine}. Kang et al. transformed the outputs of GSCMs (i.e., path and cluster parameters) into image representations and employed generative neural networks (GNN) to learn their distributions, enabling rapid generation of GSCM samples consistent with specific scene geometries \cite{kang2024geometry}. Meanwhile, the Channel Knowledge Map (CKM) approach has gained increasing attention. By leveraging interpolation techniques or deep learning methods, CKM constructs knowledge maps from limited measurement data to establish the correspondence between environmental information and channel characteristics, thereby providing a new means for channel prediction and inference in complex urban environments \cite{zeng2024tutorial,xu2024much,wu2023environment}. Although these studies address certain limitations of conventional models, most are predominantly data-driven and thus struggle to explain the physical mechanisms of signal propagation in urban canyon scenarios. In addition, their reliance on data augmentation for generalization often results in insufficient robustness and limited adaptability in highly complex urban environments. To the best of our knowledge, there is no model for inferring about the channel by relying only on environmental parameters.
 % Li et al. proposed the Digital Twin Online Channel Modeling (DTOCM) framework, which emphasizes continuous sensing, real-time updating, and integration with network optimization for site-specific online channel modeling \cite{li2025digital}
 
In this paper, we propose a site-specific channel inference model based on environmental parameters and parameterize it using large-scale measurement data collected in urban canyon scenarios. The main contributions are summarized as follows:

\begin{itemize}
  \item We propose the channel inference model, which establishes a mapping between the canyon width and the statistical characteristics of MPCs. By incorporating environmental parameters as input, the model is able to infer and generate the corresponding channel samples.
  
  \item We conducted extensive measurements in a typical urban canyon environment and extracted MPCs and developed statistical distribution models associated with canyon width, thereby enabling the parameterization of the proposed model.
  
  \item We demonstrated the application of the proposed model in map-based channel modeling tasks and validated its performance by comparing it with measurement data collected from different streets. The results show that the model can accurately capture the channel characteristics in lightly built urban environments.
\end{itemize}
The remainder of this paper is organized as follows: Section II outlines the proposed model. Section III presents the channel measurements. Section IV details the data pre-processing and modeling. Section V details the model implementation and validation. Section VI concludes the paper. 
\section{CHANNEL INFERENCE MODEL}%%%第二章
In urban canyon environments, signal propagation is strongly influenced by the distribution and placement of buildings. These site-specific environmental parameters determine the generation mechanisms and spatial distribution of multipath components, thereby affecting key channel characteristics such as power, delay, and angle of arrival. Therefore, we propose a site-specific channel inference model based on environmental parameters, as illustrated in Fig. \ref{fig:systemv2}. 

The channel inference model infers the corresponding channel from the input environmental parameters, thereby establishing a mapping from environment to channel characteristics. Specifically, the environmental parameters refer to the set of distances $\mathbf{D}$ of clusters relative to the propagation paths, expressed as:
\begin{equation}
\mathbf{D}=\left\{d_{l, 1}, d_{l, 2}, \cdots, d_{l, m}, d_{r, 1}, d_{r, 2}, \cdots, d_{r, n}\right\}
\end{equation}
where $d_{l,m}$ and $d_{r,n}$ denote the distances from the $m$-th cluster on the left side and the $n$-th cluster on the right side to the propagation path, respectively. To avoid notational ambiguity in the subsequent statistical modeling, we use the scalar $D$ exclusively to denote the effective one-sided canyon width (i.e., the lateral distance from the RX to the dominant building surfaces on the specific side, $D \in \{D_l, D_r\}$). To eliminate spatial ambiguity, "left" and "right" in this model are defined relative to the forward driving direction of the receiver vehicle along the longitudinal axis of the street. This distinction is crucial because the typical off-center topology in vehicular scenarios (e.g., a transmitter parked on one side and a receiver driving in a specific lane) introduces an inherent geometric asymmetry, making the signal interaction mechanisms and scattering contributions from the left and right sides statistically distinct. During signal propagation, the presence of clusters generates distinct discrete MPCs. Each MPC represents a set of parameters for a propagation path, including power, delay, azimuth angle of arrival (AoA), elevation angle of arrival (EoA), and phase, which are specifically expressed as:
\begin{equation}
X=\{\beta, \tau, \theta, \varphi, \psi\}
\end{equation}
where $\beta$, $\tau$, $\theta$, $\phi$, and $\psi$ denote the power, delay, AoA, EoA, and phase, respectively.

Different values of $d_l$ and $d_r$ affect the MPCs formed by the clusters. Therefore, the core idea of channel inference model is to uncover the implicit mapping $h_X$ between $d_l$ and the statistical characteristics of the channel parameters, thereby inferring and obtaining the statistical features and multipath distributions of the channel under arbitrary environmental layouts. Given a set of MPCs and $D$, $h_X$ can be expressed as:
%%%%%%%系统框图%%%%%%%%%%%%%
\begin{figure}[t]
    \centering
    \includegraphics[width=\linewidth]{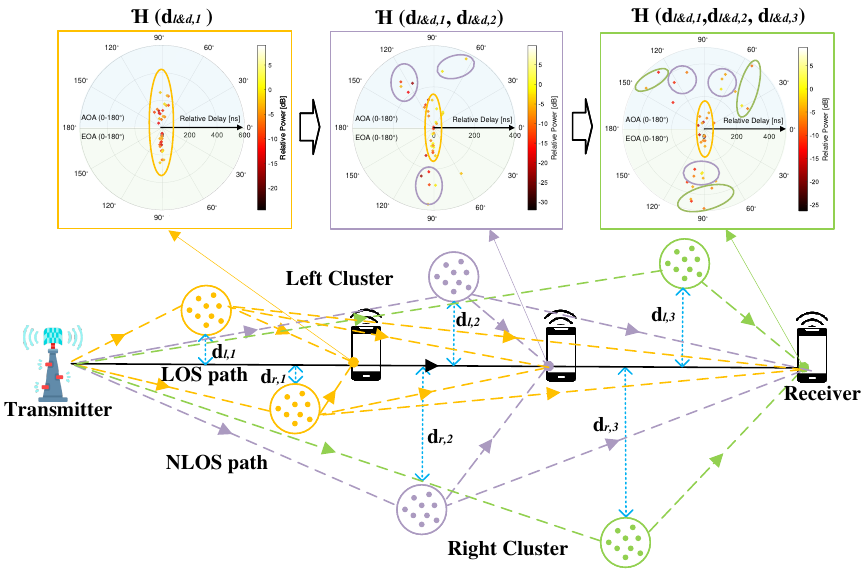}
    \caption{Schematic diagram of the proposed site-specific channel inference model.}
    \label{fig:systemv2}
\end{figure}
%%%%%%%%%%%%%%%%%%%%%%%%%%
\begin{equation}
X \mid D \sim F_X\big(\psi_X(D)\big), \quad \psi_X(D) = h_X(D; \eta_X)
\end{equation}
where $F_X(\cdot)$ denotes any latent distribution of $X$ (e.g., $N$ for Normal distribution, $L$ for Laplace distribution, $SL$ for Single-sided Laplace distribution, $E$ for Exponential distribution, etc.); $\psi_X(D)$ represents the parameter vector of this distribution (e.g., mean, variance, scale parameter, and shape parameter, etc.); $h_X$ denotes the mapping function from $D$ to $\psi_X(D)$ (e.g., linear, log-linear, piecewise, or monotonic functions, etc.); and $\eta_X$ is the set of hyperparameters to be calibrated, which can be fitted on measured or simulated data using methods such as least squares (LS), maximum likelihood estimation (MLE), or Bayesian inference (Bayes), etc.
%%%%%%%测量系统%%%%%%%%%%%%%
\begin{figure*}[t]
    \centering
    \includegraphics[width=\linewidth]{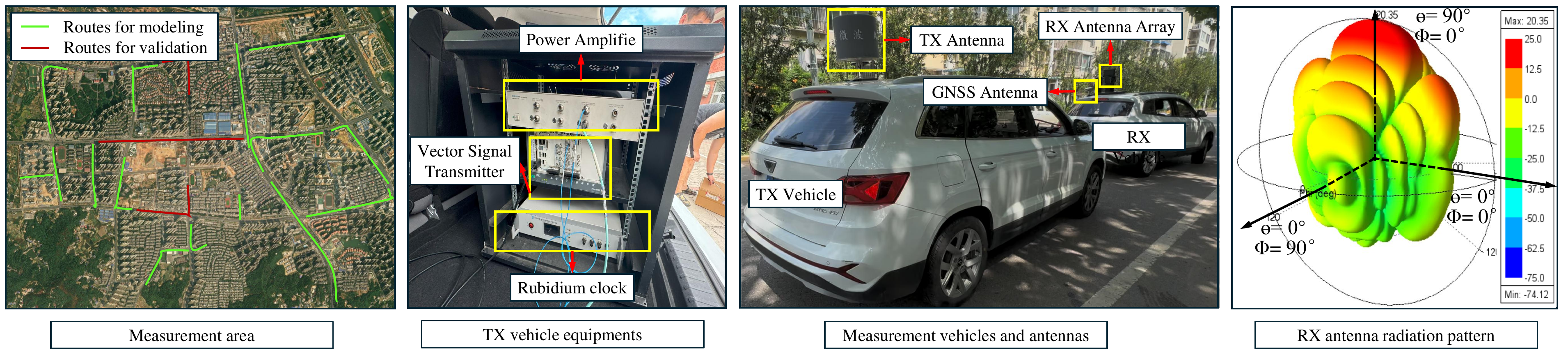}
    \caption{Measurement area, measurement system architecture and key equipment.}
    \label{fig:measurement}
\end{figure*}
%%%%%%%%%%%%%%%%%%%%%%%%%%
%%%%%%%测量路线%%%%%%%%%%%%%
\begin{figure}[t]
    \centering
    \includegraphics[width=\linewidth]{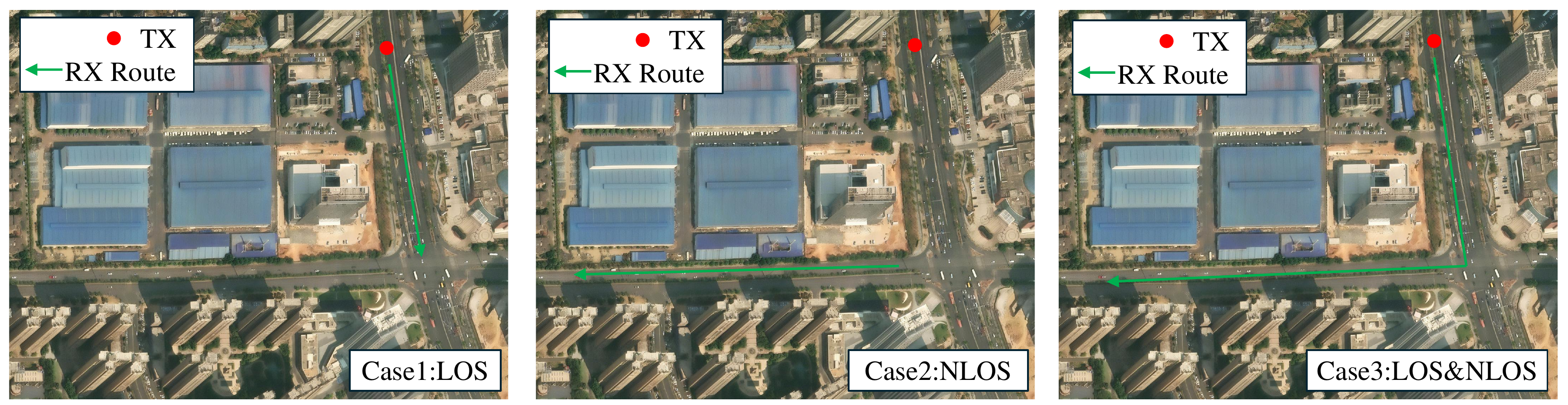}
    \caption{Measurement Scenarios.}
    \label{fig:route}
\end{figure}
%%%%%%%%%%%%%%%%%%%%%%%%%%

For time-varying channels, the geometric relationships between environmental scatterers and the transceiver change over time, causing the parameters $X$ of the MPCs generated by these scatterers to exhibit temporal evolution. In addition, each MPC has a finite lifecycle (birth–death). To accurately describe this time-varying process, we extend the environmental parameters $D$ set to a time-dependent function $D(t)$ and introduce the state space $s_{k,i}(t) \in \{0,1\}$ for each MPC: when $s_{k,i}(t)=1$, the $i$-th subpath of the $k$-th cluster exists at time $t$; otherwise, it does not exist. The number of MPCs in cluster $S_k$ at time $t_i$ can be expressed as:
\begin{equation}
L(t_i,k) = L(t_i,i,k) + \sum_{j=1}^{i-1} L(t_j,i,k), \quad (0 < j < i)
\end{equation}
where $L(t_i,i,k)$ denotes the number of newly generated MPCs at $t_i$ that exist at $t_i$, and $\sum_{j=1}^{i-1} L(t_j,i,k)$ represents the sum of MPCs that were generated at earlier times and survive until $t_i$.

For a given environmental layout, the channel impulse response (CIR) at time $t$ can be reconstructed by superimposing the contributions of all MPCs, based on the temporal evolution of clusters and their subpaths along with the corresponding multipath parameters. This can be expressed as:
\begin{equation}
\begin{split}
h(t) &= \sum_{k=1}^{K(t)} \sum_{i=1}^{L(t,k)} 
    s_{k,i}(t)\ \beta_{k,i}\big(D(t)\big) 
    e^{\,j \psi_{k,i}\big(D(t)\big)} \\
    &\quad \delta\big(t - \tau_{k,i}\big(D(t)\big)\big) 
    a\Big(\theta_{k,i}\big(D(t)\big), \phi_{k,i}\big(D(t)\big)\Big)
\end{split}
\end{equation}

We have designed a visualization of the MPCs, as illustrated in Fig. \ref{fig:systemv2}. Within a single snapshot, the channel parameters are plotted in a polar coordinate system, where the axes are divided into two semicircles: The upper semicircle represents the AoA, and the lower semicircle represents the EoA. The radius corresponds to the delay, and the color represents the power. This visualization method provides an intuitive representation of how the channel characteristics vary with $D(t)$.

\section{CHANNEL MEASUREMENTS}%%%第三章
In this section, we describe the previous measurement campaigns conducted for channel inference model. These measurements provide the foundation for developing the channel model, with the model structure based on the collected data and the model parameters extracted from it.
\subsection{Measurement Setup}%%%测量设备
% %%%%%%%%%%%%%%%%%%%%%%%
\begin{table}[t]
\centering
\caption{Configurations of measurement system}
\begin{tabular}{lc}
\toprule
\textbf{Parameters} & \textbf{Value} \\
\midrule
Center frequency           & 5.8\,GHz \\
Bandwidth                  & 30\,MHz \\
Transmit power             & 45\,dBm \\
Sounding signal            & Multi-carrier signal \\
Number of frequency points & 1024 \\
Transmitter antenna        & Omnidirectional antenna \\
Receiver antenna           & 4$\times$8 planar array \\

\bottomrule
\end{tabular}
\label{tab:config}
\end{table}
We conducted measurement campaigns using the equipment shown in Fig. \ref{fig:measurement}. The measurement system consists of transmitting (Tx) and receiving (Rx) subsystems integrated into vehicles, with the core components being a vector signal generator (VSG) and a vector signal analyzer (VSA). Specifically, the VSG and VSA are NI PXIe-5673 and NI PXIe-5663, respectively. The sounding signal is a broadband multi-carrier signal with 513 subcarriers over a bandwidth of 30 MHz.
%%%%%%%数据处理框图%%%%%%%%%%%%%
\begin{figure}[t]
    \centering
    \includegraphics[width=\linewidth]{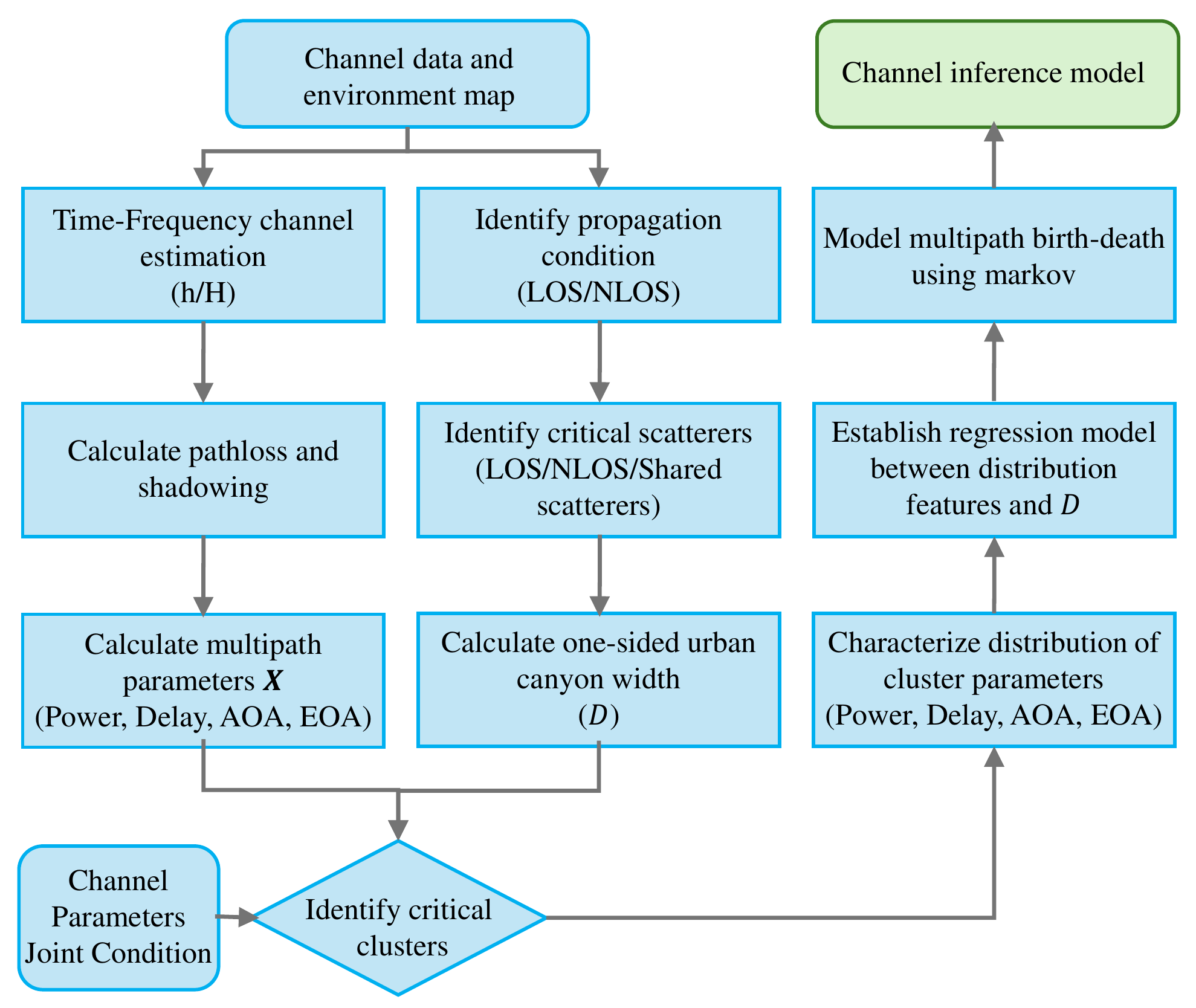}
    \caption{Procedure of establishing site-specific channel inference model.}
    \label{fig:datapre}
\end{figure}
%%%%%%%%%%%%%%%%%%%%%%%%%%

The TX antenna is an omnidirectional single element antenna, while the RX antenna is a 4×8 array. Both TX and RX antennas are mounted on the vehicle roofs at a height of approximately 1.8 m. The Z axis of the RX antenna array is oriented opposite to the vehicle’s forward direction. To obtain accurate angle information using the antenna array, the radiation patterns of both the TX antenna and RX antenna arrays were measured in an anechoic chamber. The right side of Fig. \ref{fig:measurement} shows the 3D radiation pattern of one RX element. Each element of the RX antenna array is connected to a vector signal analyzer through an electronic switch. For precise time synchronization, two rubidium clocks disciplined by Global Navigation Satellite System (GNSS) signals are employed. These clocks also provide real-time longitude and latitude coordinates, enabling accurate positioning of both the Tx and Rx.

Table I summarizes the detailed configurations of the measurement system. The measurement bandwidth is 30 MHz, which results in a delay resolution of 33.3 ns. This delay resolution means that only MPCs with propagation distance differences exceeding 10 m can be distinguished in the delay domain. For sub-6 GHz vehicular communications, the available bandwidth is generally 20–30 MHz, which is similar to the measurement configuration in this article. The acquisition rate of channel snapshots is 45 snapshots/s when measured.
\subsection{Measurement Campaign}%%%测量活动
%%%%%%%数据分析%%%%%%%%%%%%%
\begin{figure}[t]
\centering
\subfigure[]{\includegraphics[width=2.1in]{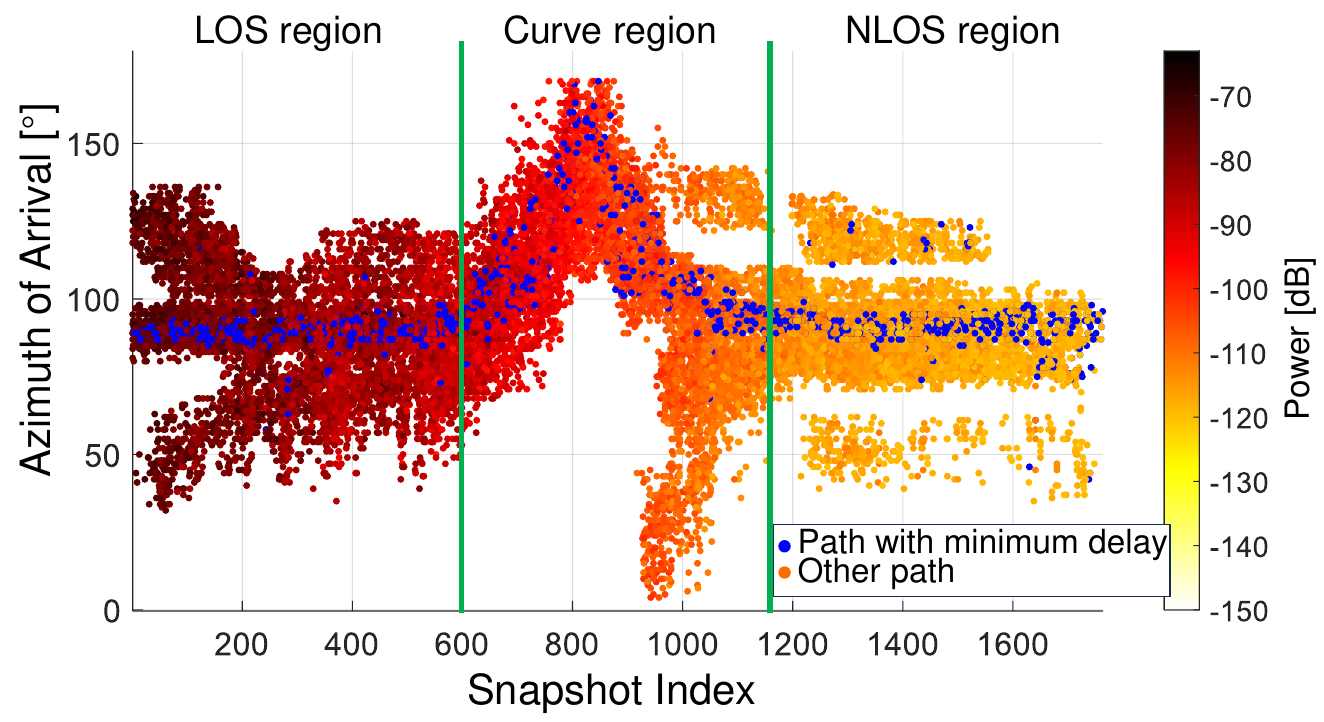}}
\subfigure[]{\includegraphics[width=1.0in]{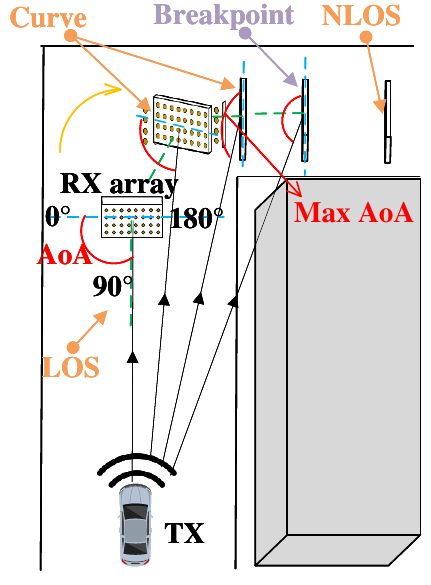}}
\caption{Data preprocessing and propagation patterns analysis. (a) MPCs extracted by SAGE. (b) Schematic diagram of AoA variation during turning maneuvers.}
\label{fig:dataanly}
\end{figure}
%%%%%%%%%%%%%%%%%%%%%%%%%%
The measurements were conducted in Changsha, China. As shown in the aerial photographs in Fig. \ref{fig:measurement}, the measurement area covers major roads within a 4 km × 3 km urban district. The streets under measurement are flanked by densely packed buildings reaching several tens of meters in height, forming a typical urban canyon environment. As illustrated in Fig. \ref{fig:route}, the measurement routes comprise three cases:
\begin{itemize}
  \item \textbf{Case1: LOS scenarios.} TX and RX vehicles were on the same road, with the RX vehicle positioned ahead of the TX vehicle. The TX vehicle was parked at the roadside, while the RX vehicle maintained an average speed of 30 km/h to ensure consistency in data collection.
  
  \item \textbf{Case2: NLOS scenarios.} RX vehicle traveled on a road adjacent to and approximately perpendicular to the one where the TX was located. TX vehicles parked on the side of the road at the beginning of the process, and the RX maintained the same speed as in the LOS cases.
  
  \item \textbf{Case3: LOS\&NLOS scenarios.} TX and RX vehicles were initially on the same road, after which the RX vehicle turned onto another road. TX vehicles parked on the side of the road at the beginning of the process, and the RX maintained the same speed as in the LOS cases.
\end{itemize}

Measurements were conducted in the absence of other nearby vehicles to ensure that additional vehicular movements did not influence the results used for analysis.

\section{DATA PROCESSING AND MODELING}%%%第四章
This section describes how we construct the proposed model based on measurement data, with the data processing workflow illustrated in Fig. \ref{fig:datapre}. MPCs are extracted using SAGE \cite{fessler2002space,matthaiou2007characterization}, a widely used high-resolution parameter estimation (HRPE) algorithm based on the expectation-maximization (EM) framework. We perform HRPE for each snapshot and then, based on propagation characteristics, we derive the constraint relationships between the channel parameters and the environmental parameters. Using these relationships, we identify the MPCs corresponding to specific scatterers. We then provide a detailed explanation of how these MPCs are employed to construct the model proposed in Section II.
%%%%%%%传播框图%%%%%%%%%%%%%
\begin{figure}[t]
    \centering
    \includegraphics[width=\linewidth]{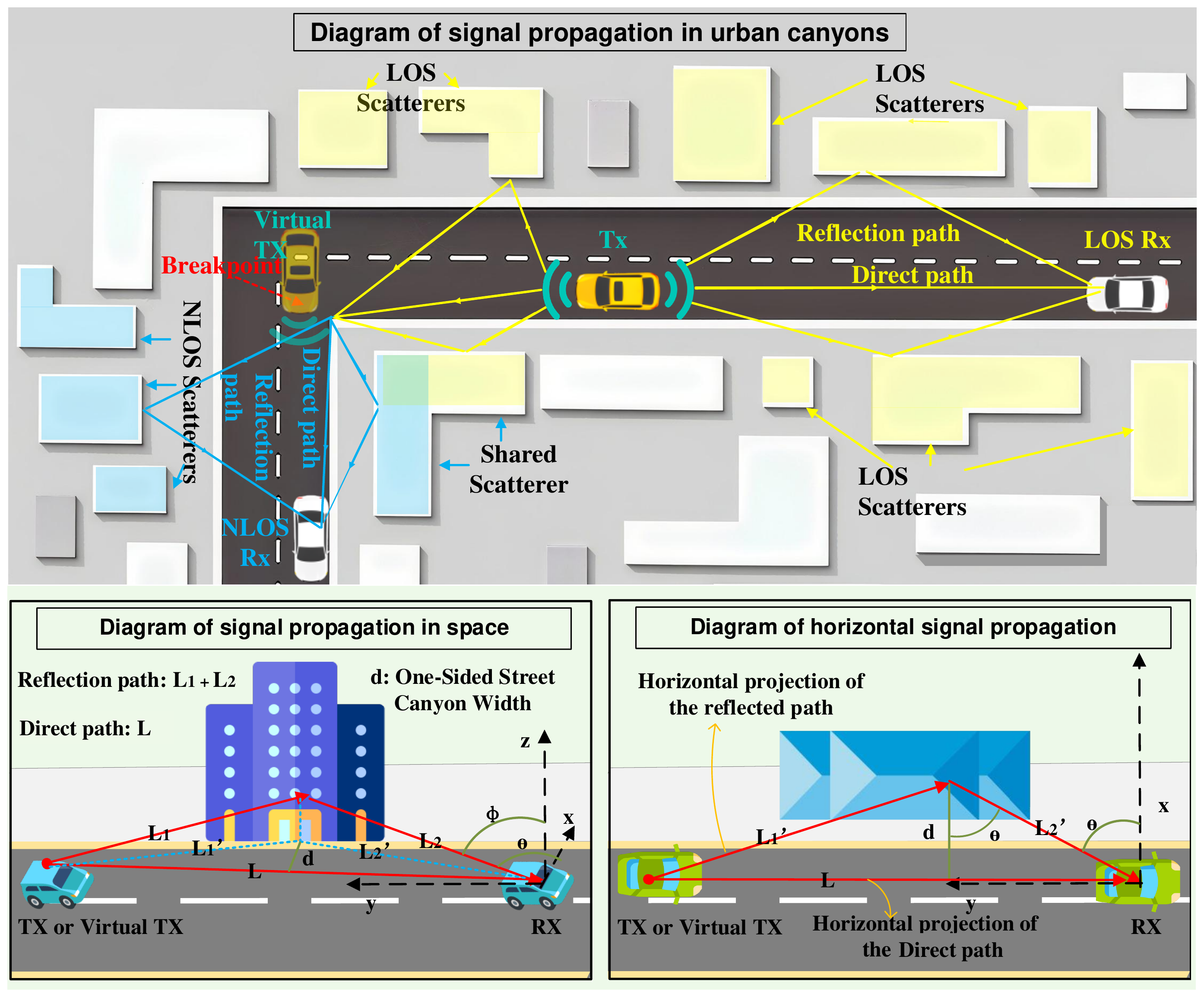}
    \caption{Diagram of signal propagation in urban canyons.}
    \label{fig:urban}
\end{figure}
%%%%%%%%%%%%%%%%%%%%%%%%%%
\subsection{Signal propagation patterns in urban canyons}%%%传播规律
The propagation patterns of signals in urban canyon form the foundation for establishing the proposed model. Therefore, we analyze a set of measurement data from Case 3. The MPCs extracted by SAGE are shown in Fig. \ref{fig:dataanly}(a), where, in each snapshot, the path with the minimum delay is highlighted in blue. Within the LOS region, the minimum-delay path corresponds to the direct path, with the AoA distributed around 90°. Distinct clusters can be observed on either side of the direct path, representing multipath components generated by scatterers located on both sides of the canyon.

In the curve region, as the normal direction of the RX antenna array gradually rotates, the AoA continuously increases. When the RX vehicle completes the turn, the AoA reaches its maximum. Subsequently, as the RX vehicle travels perpendicular to the TX vehicle, the AoA gradually decreases until, at the breakpoint, the direct path disappears entirely and the RX vehicle enters the NLOS region. Fig. \ref{fig:dataanly}(b) illustrates how the AoA varies with the motion of the RX vehicle.
%%%%%%%%%%%%%%多径识别%%%%%%%%%%%%
\begin{figure}[t]
    \centering
    \includegraphics[width=0.95\linewidth]{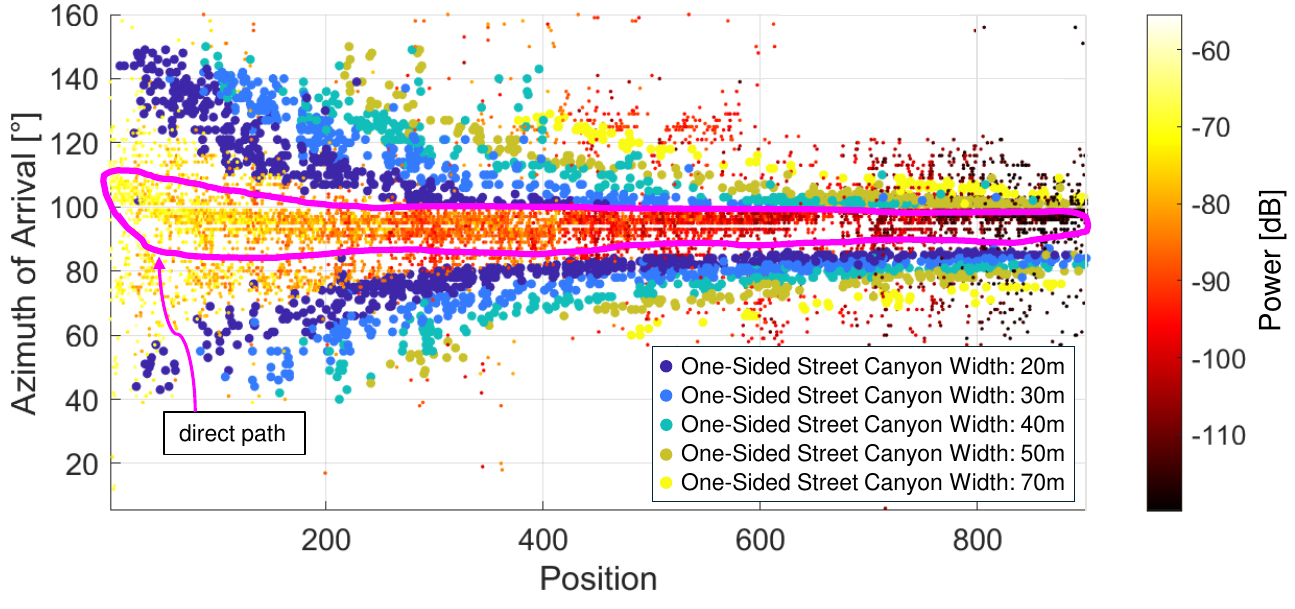}
    \caption{MPCs identification results.}
    \label{fig:Idenfiy}
\end{figure}
%%%%%%%%%%%%%%%%%%%%%%%%%%%%%%

It should be noted that during the RX vehicle’s turn, the MPCs consist solely of the direct path. After the RX completes the turn and before reaching the breakpoint, multipath components generated by scatterers reappear. Furthermore, within the NLOS region, the formation pattern of AoA is similar to that in the LOS region: all minimum-delay paths form a cluster resembling the direct path, with distinct scatterer clusters on either side. However, the power of these multipaths exhibits abrupt changes at the breakpoint, indicating that diffraction and bending of the signal occur there. After the RX enters the NLOS region, it can be equivalently considered that a virtual TX located at the breakpoint transmits signals to the RX. These signals also propagate via scatterers on both sides of the canyon within the NLOS region, generating the observed MPCs.

Based on the above analysis, we define a schematic representation of site-specific propagation in an urban canyon, as shown in Fig. \ref{fig:urban}, which illustrates the propagation under LOS and NLOS conditions and categorizes scatterers into three types: LOS scatters, NLOS scatters, and shared scatters. 
% Since the measurements were conducted in winter, the influence of foliage is negligible, and only building scatterers on both sides of the road are considered. 

Fig. \ref{fig:urban} analyzes the impact of a single scatterer on the MPCs and defines the straight-line distance between the building and the transceiver as the one-sided canyon width, which has the same meaning as $D$ defined in Section II. It should be noted that this definition differs from the conventional canyon width, as scatterers on the left and right sides generate MPCs with different characteristics. In this work, all references to the canyon width denote the one-sided canyon width. 
%%%%%%%%%%%%%%反演位置%%%%%%%%%%%%
\begin{figure}[t]
    \centering
    \includegraphics[width=0.95\linewidth]{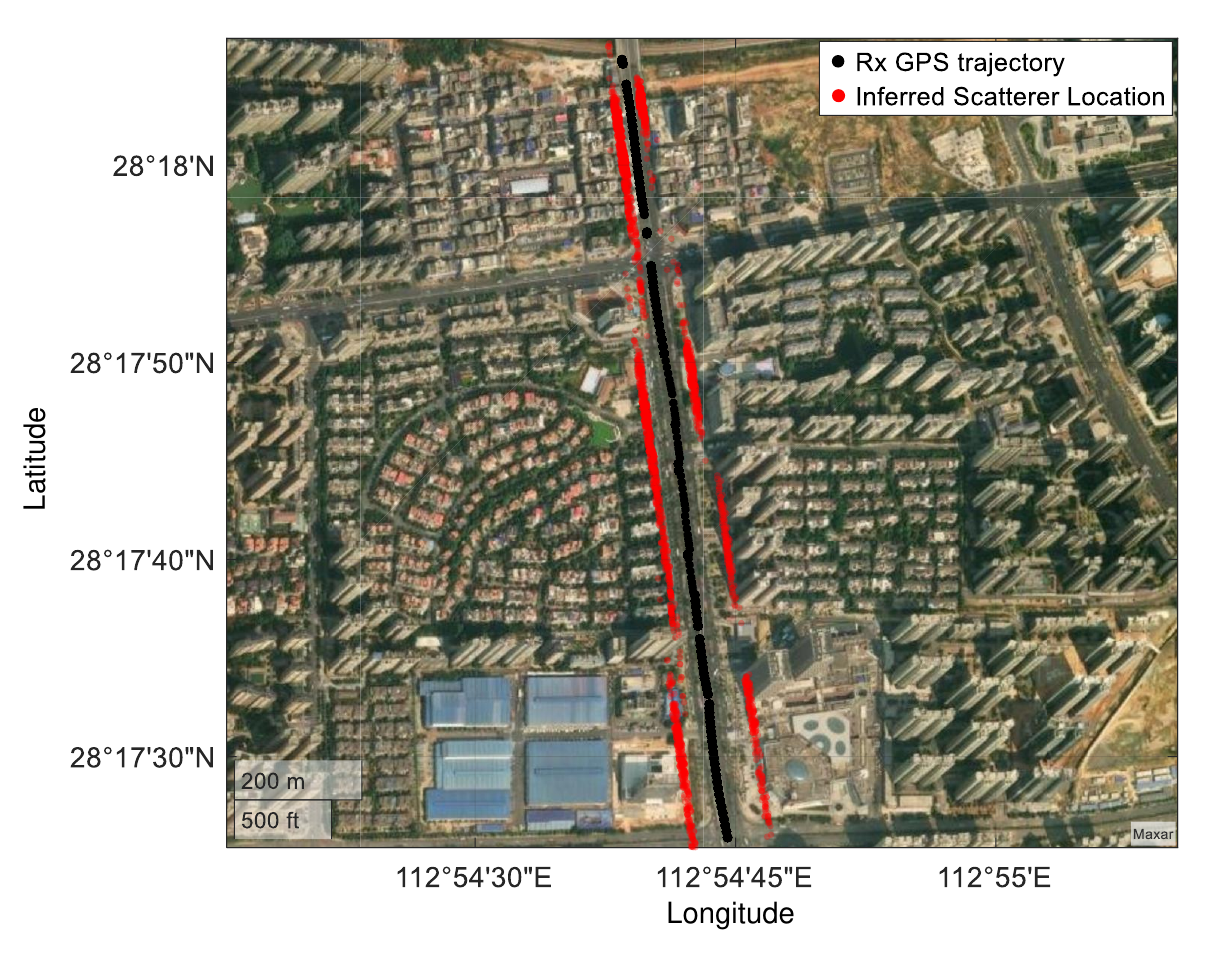}
    \caption{Scatterers reconstructed from MPCs.}
    \label{fig:Interred}
\end{figure}
%%%%%%%%%%%%%%%%%%%%%%%%%%%%%%

Based on the geometric propagation relationships illustrated in Fig. \ref{fig:urban}, at any given time, the constraint relationship between the MPCs and $d$ can be expressed as (6),
\begin{figure*}[t]
\hrulefill % 顶部分割线（可选）
\vspace{0.5em}

% 方法 A：在公式上方直接添加描述性文字（最推荐，符合大多数IEEE期刊风格）
\noindent \textit{1) MPCs Identification under Geometric Constraints:}
\begin{equation}
\label{eq:mpc_extraction}
\frac{D_i}{c} 
\Bigg( 
\sqrt{
1 + 
\Bigg( \frac{\tau_{i,\text{direct}} c}{D_i}
\sin \varphi_{i,\text{direct}} - \tan \theta_{i,\text{reflect}} 
\Bigg)^2 
+ \frac{1}{\cos^2 \theta_{i,\text{reflect}}} 
} 
\Bigg)
- \tau_{i,\text{reflect}} \sin \varphi_{i,\text{reflect}}
\le \delta \tag{6}
\end{equation}

\vspace{1em}
\noindent \textit{2) Scatterer Position Reconstruction ($D_i, L_i$):}
\begin{equation}
\label{eq:reconstruct_d}
D_i = \frac{\tau_{i,\text{reflect}}  \sin \varphi_{i,\text{reflect}}  c}{\sqrt{ 1 + \left( \frac{\tau_{i,\text{direct}}  c}{D_i}  \sin \varphi_{i,\text{direct}} - \tan \theta_{i,\text{reflect}} \right)^2 + \frac{1}{\cos^2 \theta_{i,\text{reflect}}} }} \tag{7}
\end{equation}
\begin{equation}
\label{eq:reconstruct_l}
L_i = \tau_{i,\text{direct}} \cdot c - \Big| \tan(\theta_{i,\text{reflect}}) \cdot \frac{D_i}{2} \cdot \sin(\varphi_{i,\text{reflect}}) \cdot \cos(\theta_{i,\text{reflect}}) \Big| \tag{8}
\end{equation}

\vspace{0.5em}
\hrulefill % 底部分割线
\end{figure*}
where $d_i$ denotes the one-sided canyon width corresponding to the $i$-th snapshot; $\tau_{i,\text{direct}}$ and $\tau_{i,\text{reflect}}$ represent the delays of the direct and reflected paths, respectively; $\theta_{i,\text{reflect}}$ denotes the AoA of the reflected path; $\phi_{i,\text{direct}}$ and $\phi_{i,\text{reflect}}$ represent the EoA of the direct and reflected paths, respectively; and $c$ is the speed of light. The error threshold $\delta$ accounts for the electromagnetic effects of practical propagation environments, such as the dielectric properties of buildings. In this work, $\delta=3.3\times10^{-8}$, corresponding to the delay resolutio of the measurement equipment, indicating that this constraint relationship can distinguish MPCs generated by buildings whose widths differ by more than 10 m. Based on the analysis of signal propagation patterns presented in this section, (6) is applicable to both LOS and NLOS scenarios.
\subsection{Multipath component identification and verification}%%%多径识别和验证
Based on Section III-A, it is only necessary to determine the canyon width between the transmitter and the receiver to identify the MPCs for each snapshot using (6). This operation does not depend on the precise distribution of scatterers but only on the effective widths. The threshold $\delta$ in (6) acts as a geometric gate. In scenarios where the TX and RX are far apart, any residual errors introduced by approximating a multi-bounce path as an equivalent single-bounce path are absorbed into the statistical scale parameters (e.g., the delay spread of the cluster). By doing so, the model implicitly accounts for the increased complexity of the propagation environment at large distances without requiring an exhaustive deterministic description of every reflection surface.

We selected a set of measurement data from an LOS case for the identification of MPCs, as shown in Fig. \ref{fig:Idenfiy}. The multipaths identified according to different canyon widths naturally form clusters, which are highlighted in different colors in Fig. \ref{fig:Idenfiy}. It can be observed that as the width increases, the AoA of the corresponding clusters deviates further from the direct path, consistent with the physical propagation principles.

Using the delay, AoA and EoA of the reflected path along with the delay of the corresponding direct path, the spatial position of the reflection point can be reconstructed by (7) and (8), where $D$ and $L$ denote the reconstructed one-sided canyon width and the distance of the reflection point from TX, respectively, while the other parameters are defined as in (6).

By converting the planar coordinates ($D$ and $L$) to longitude and latitude, the corresponding scatterer positions can be inferred on the aerial photograph, as shown in Fig. \ref{fig:Interred}. The black points represent the RX trajectory, and the red points indicate the scatterer positions inferred from the MPCs. These inferred scatterers align well with the heights of buildings on both sides of the road, demonstrating the validity of (6). It should be noted that the influence of a canyon on MPCs is not confined to its immediate location. For example, a canyon width of 20 m appears near the starting point, yet the MPCs it generates persist over a much longer distance range.
\subsection{Channel Inference Model}%%%建模
%%%%%%%参数分析20图%%%%%%%%%%%%%
\begin{figure*}[t]
    \centering
    \includegraphics[width=\linewidth]{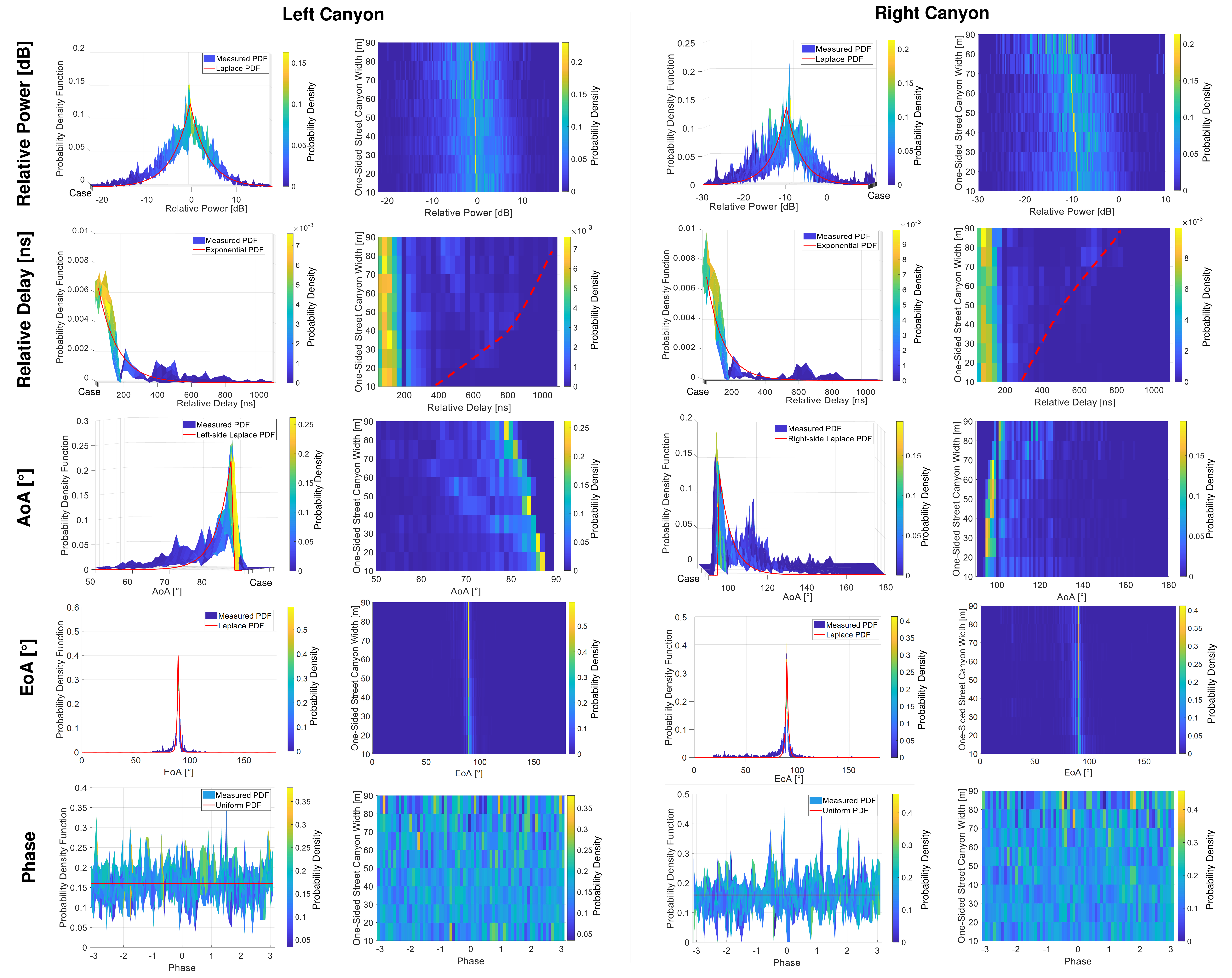}
    \caption{Statistical distributions of MPCs parameters for various urban canyon widths. As $D$ increases, the relative power mean decreases due to higher path loss, while the relative delay mean and spread increase; the AoA distribution peak shifts according to geometric parallax, whereas EoA and phase remain largely insensitive to $D$.}
    \label{fig:distutions}
\end{figure*}
%%%%%%%%%%%%%%%%%%%%%%%%%%
\textit{1) MPCs statistical distribution}: As described in Section II, the core idea of the proposed model is to construct the mapping function $h_X$ in (3), which represents the relationship between canyon widths $D$ and the statistical characteristics of the MPCs. We adopt a two-step approach: first, we determine the statistical distribution $F_X$ of the MPCs and the statistical features $\psi_X$ influenced by $D$; second, we construct the mapping function $h_X$ between the distribution features and the canyon widths, thereby obtaining the distribution of MPCs as a function of $D$, expressed as $F_X(h_X(D;\eta_X))$.

In the proposed modeling approach, the MPCs are explicitly partitioned into left-side and right-side clusters. It should be emphasized that electromagnetic waves naturally do not distinguish between "left" and "right", rather, this bipartite modeling strategy is designed to capture the inherent geometric asymmetry of practical urban vehicular communications.In typical V2I or V2V scenarios (as configured in our measurements), the TX is not located at the exact geometric center of the canyon, but is typically parked on one side of the road (e.g., the right curb). Similarly, following specific traffic rules, the RX vehicle travels along a specific lane, resulting in an off-center trajectory. This off-center topology means that the relative TX-scatterer-RX signal interaction geometries are inherently asymmetric between the two sides of the street. Consequently, even if a building on the left and a building on the right share the same absolute distance $D$ to the RX, their scattering contributions to the channel are statistically different due to their different relative positions to the TX. Therefore, modeling them separately prevents the distinct asymmetric spatial topology from being averaged.

Consequently, we identify the MPCs corresponding to different canyon widths from the measurement data and distinguish between the left and right sides of the canyon, generating five parameters: relative power, relative delay, AoA, EoA, and phase. Here, the relative power and the relative delay represent the deviations of the reflected paths from the direct path. Fig. \ref{fig:distutions} presents the statistical distributions of these parameters under different canyon widths. In the first and third columns, the widths axis is compressed to observe whether channel parameters under different widths follow the same distribution. The second and fourth columns show the Probability Distribution Functions (PDP) for different $D$. The following provides an analysis of each parameter:    
\begin{itemize}
  \item \textbf{Relative power:} For relative power, its distribution can be modeled as a Laplace distribution. As the canyon width $D$ increases, the mean $u_\beta$ gradually decreases. This is because a larger $D$ results in longer propagation paths for the reflected path relative to the direct path, leading to increased free-space loss. We select a linear regression function as $h_\beta$ for $u_\beta$, while the scale parameter $b_\beta$ remains largely consistent across discrete values and can be considered unaffected by $D$. Accordingly, the distribution function can be expressed as $F_\beta(h_\beta(D;\eta_\beta))$:
  \begin{equation}
f(\beta \mid D) = \frac{1}{2 b_\beta} \exp\Bigg(-\frac{ \big| \beta - (\alpha_\beta D + \beta_\beta^{(0)}) \big| }{b_\beta} \Bigg)
\end{equation}
  where $\{\alpha_\beta, \beta_\beta^{(0)}\} \in \eta_\beta$ is a set of hyperparameters fitted from the measurement data, and $b_\beta$ is determined from the statistical average of the measured data.
  
  \item \textbf{Relative delay:} For relative delay, its distribution can be modeled as an exponential distribution. As $D$ increases, the spread of relative delay becomes larger and the decay of the distribution slows, indicating increased delay offsets. Similarly, for larger $D$, the reflected paths require longer propagation times to complete their trajectories. Accordingly, we select a linear regression function as $h_\tau$ for the mean $u_\tau$ (in an exponential distribution, the standard deviation is equivalent to the mean, both represented as $1/\lambda$). The distribution can thus be expressed as $F_\tau(h_\tau(D;\eta_\tau))$:
  \begin{equation}
f(\tau \mid D) = \frac{1}{\alpha_\tau D + \beta_\tau^{(0)}} \exp\Bigg( - \frac{\tau}{\alpha_\tau D + \beta_\tau^{(0)}} \Bigg)
\end{equation}
  where $\{\alpha_\tau, \beta_\tau^{(0)}\} \in \eta_\tau$ is a set of hyperparameters fitted from the measurement data. Only the portion with $\tau > 0$ is considered here, because the delay of the direct path is nearly the minimum in each snapshot, making all relative delays essentially greater than zero.
  
  \item \textbf{AoA:} For the AoA, its distribution can be modeled as a single-sided Laplace distribution. Based on the orientation of the array antenna shown in Figure 5(b), the AoA range for the left canyon is 0–90°, while that for the right canyon is 90–180°. If the AoA of the direct path (approximately around 90°) is also included, the global AoA approximately follows a Laplace distribution, consistent with conventional modeling practices.

  For the AoA of the left canyon, we assume that the single-sided distribution can be mirrored to the right side to construct a full Laplace distribution. The mean $u_{\theta,l}$ of this distribution is taken as the maximum AoA observed in the measurement data, allowing the mapping between $\theta_\text{max}$ and $D$ to be fitted via linear regression as $h_{\theta,l}$. Therefore, when generating the AoA for the left canyon, only the single-sided range needs to be considered. The mapping function $h_{\theta,r}$ for the right canyon can be fitted using the same approach. The distribution function can thus be expressed as $F_\theta(h_\theta(D;\eta_\theta))$:
  \begin{equation}
  f(\theta \mid D) = 
\begin{cases} 
\frac{1}{b_{\theta,l}} \exp\left( \frac{\theta - (\alpha_{\theta,l} D + \beta_{\theta,l}^{(0)})}{b_{\theta,l}} \right)&\text{,left} \\
\frac{1}{b_{\theta,r}} \exp\left( -\frac{\theta - (\alpha_{\theta,r} D + \beta_{\theta,r}^{(0)})}{b_{\theta,r}} \right)&\text{,right}
\end{cases}
  \end{equation}
  %%%%%%%%%%%%%%PL%%%%%%%%%%%%
\begin{figure}[t]
    \centering
    \includegraphics[width=0.95\linewidth]{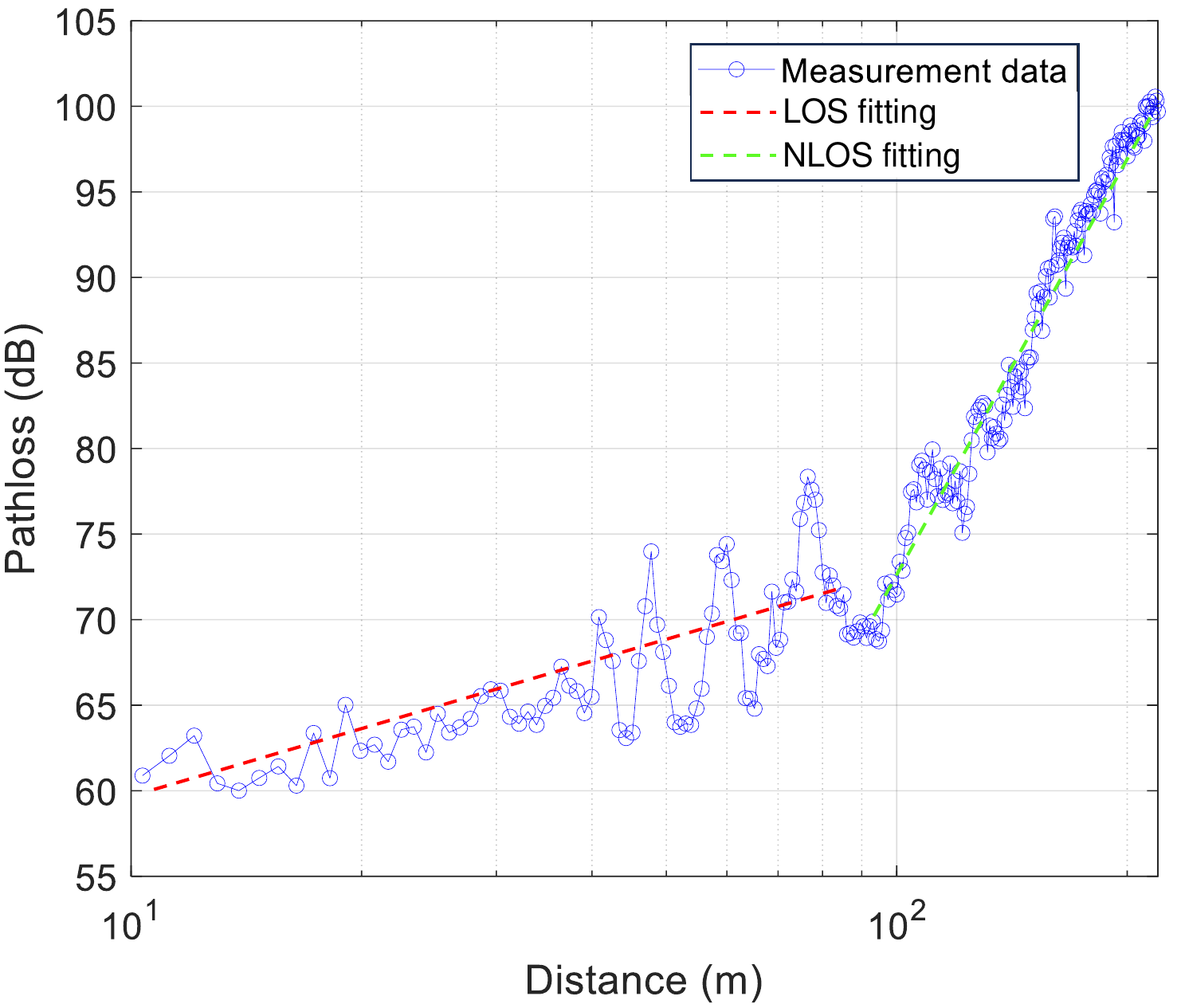}
    \caption{Log-distance fitting of measured path loss.}
    \label{fig:PL}
\end{figure}
%%%%%%%%%%%%%%%%%%%%%%%%%%%%%%
  \item \textbf{EoA:} For the EoA, its distribution can be modeled as a Laplace distribution. There is no significant difference between the EoA of the left and right sides of the canyon, and the distribution characteristics show no obvious dependence on $D$. The distribution function can thus be expressed as $F_\phi$:
  \begin{equation}
f(\phi \mid u_\phi, b_\phi) = \frac{1}{2 b_\phi} \exp\Bigg( - \frac{|\phi - u_\phi|}{b_\phi} \Bigg)
  \end{equation}
  where $u_\phi$ and $b_\phi$ can be determined from the statistical analysis of the measured data.
  
  \item \textbf{Phase:} For the phase, its distribution can be modeled as a uniform distribution. Similar to the EoA, the distribution of phase is also independent of $D$. The distribution function can thus be expressed as $F_\psi$:
  \begin{equation}
f(\psi) = \frac{1}{2\pi}, \quad \psi \in [-\pi, \pi)
\end{equation}
\end{itemize}
All the aforementioned $\eta_X$ parameters are listed in Table II. The necessity of this asymmetric modeling is strongly supported by the extracted parameters presented in Table II. As observed, the mapping coefficients (i.e., slopes $\alpha$ and intercepts $\beta^{(0)}$) for the left and right clusters exhibit notable numerical differences. For instance, the slope governing the relationship between the canyon width and the relative delay mean ($\alpha_{\tau}$) for the right side (1.0764) is significantly higher than that for the left side (0.5533). Similarly, the relative power intercepts ($\beta_{\beta}^{(0)}$) show a substantial gap between the right (-8.9410 dB) and left (-0.0733 dB) sides, indicating that the MPCs from the right side experience much higher power attenuation relative to the direct path.

This empirical non-symmetry can be directly attributed to the geometric structure of the off-center transceiver topology. Because the TX is parked on a specific roadside and the RX travels in a specific lane (e.g., closer to the right street boundary in this measurement), the dominant propagation mechanisms interact with the spatial environment entirely differently on each side. Specifically, the shorter distance to the right-side buildings forces the signal paths to graze the near-road environment, making them highly susceptible to direct obstruction and scattering by dense roadside obstacles such as trees, streetlights, or parked vehicles. This severe geometry-induced blockage explains why the power of right-side clusters is generally lower than that of the left-side clusters. In contrast, MPCs originating from the far side (the left side) traverse the open vehicular lanes, experiencing less severe attenuation from roadside vegetation and obstacles. Consequently, this asymmetric fitting phenomenon also demonstrates that the proposed model can inherently encapsulate the attenuation effects of other unspecified roadside scatterers (e.g., trees) without requiring deterministic reconstructions. If the left and right MPCs were forcibly merged, this vital spatial signature would be averaged out, deteriorating the inference accuracy.
\begin{table*}[!htbp]
\centering
\caption{CHANNEL INFERENCE MODEL PARAMETERS}
\includegraphics[width=\linewidth]{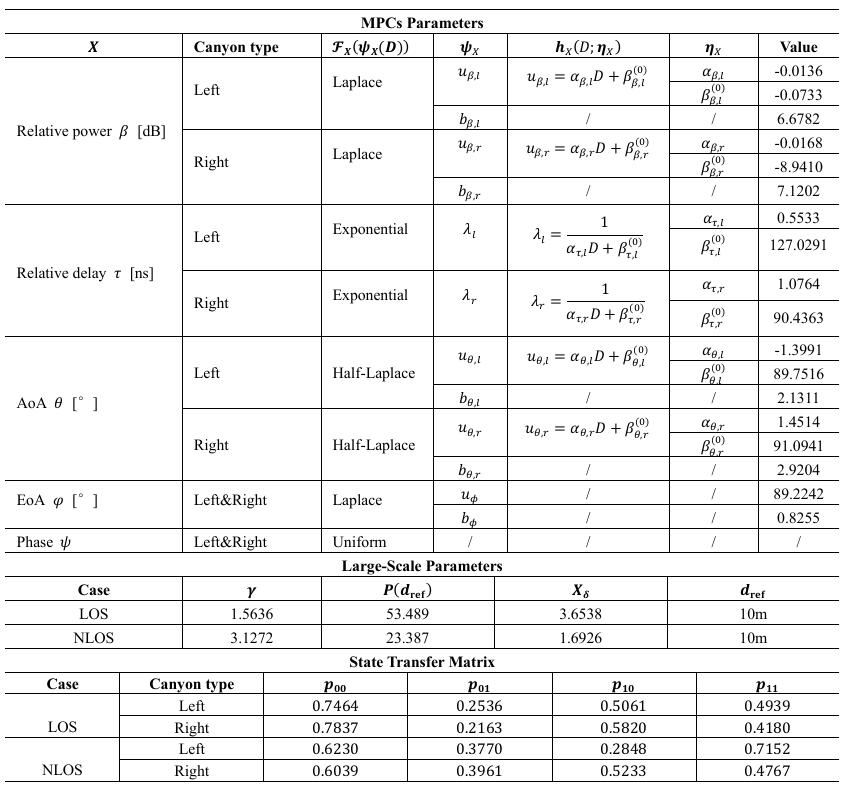}
\end{table*}

\textit{2) Large-scale parameters}: Based on the foregoing analysis and experience during data processing, the impact of canyon widths $D$ on the channel primarily manifests in the small-scale fading of the MPCs. For large-scale fading, we employ the widely used log-distance model \cite{molisch2012wireless}:
\begin{equation}
\text{PL(dB)} = P(d_{\text{ref}}) - 10 \gamma  \log\frac{d}{d_{\text{ref}}} + X_\delta
\end{equation}
where $\gamma$ is the path-loss exponent, characterizing how path loss changes with distance. $P(d_\text{ref})$ denotes the intercept of the path-loss model at the reference distance $d_\text{ref}$, and $X_\delta$ is a zero-mean Gaussian random variable describing the shadowing effect.

Specifically, we select a set of measurement data from Case 3 (LOS \& NLOS) for analysis, enabling the construction of path-loss models under both LOS and NLOS conditions. This is particularly useful for simulating the virtual TX power at the breakpoint in the NLOS model. Fig. \ref{fig:PL} presents the path-loss fitting results under LOS and NLOS conditions. It should be explicitly noted that all the large-scale parameters and the aforementioned site-specific small-scale hyperparameters $\eta_X$ listed in Table II are statistically modeled and fitted utilizing exclusively the measurement data collected from the "Routes for modeling" illustrated in Fig. 2.
  %%%%%%%ICM流程%%%%%%%%%%%%%
\begin{figure}[t]
    \centering
    \includegraphics[width=\linewidth]{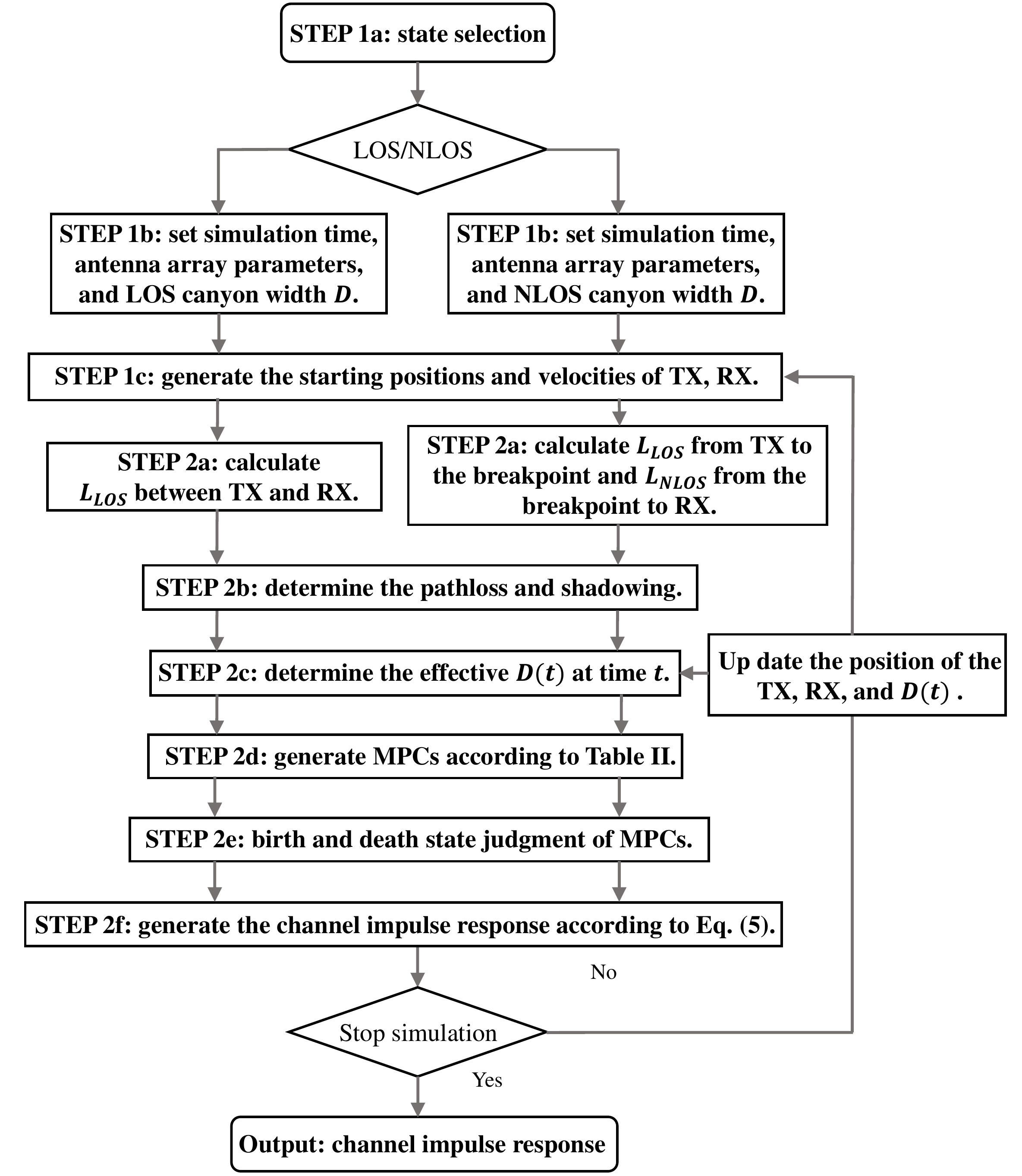}
    \caption{Flowchart of the implementation steps.}
    \label{fig:Model}
\end{figure}
%%%%%%%%%%%%%%%%%%%%%%%%%%

\textit{3) Evolution of MPCs}: It also needs to model the birth–death process of multipath components. To this end, we establish a Markov model for the identified MPCs to analyze their state transition behavior. A Markov model is a probabilistic model describing the transitions of system states over time, assuming that the state at the next time step depends only on the current state, exhibiting a memoryless property.

For each MPC, its “alive” and “dead” states within each time slot are abstracted as two states in the Markov process. According to the state space defined in Section II, $s_{k,i}(t)\in\{0,1\}$, where 0 denotes the “dead” state and 1 denotes the “alive” state. For the $k$-th cluster composed of $N$ paths, the overall cluster state is defined as the logical OR of the individual path states:
\begin{equation}
C_k(t) = \bigvee_{i=1}^{N} S_{k,i}(t)
\end{equation}
where $C_k(t)=0$ indicates that all paths are dead, and $C_k(t)=1$ indicates that at least one path is alive at time $t$.

For the $i$-th path, its state transition probability matrix can be expressed as:
\begin{equation}
P^{(i)} =
\begin{bmatrix}
p_{00}^{(i)} & p_{01}^{(i)} \\
p_{10}^{(i)} & p_{11}^{(i)}
\end{bmatrix}
\end{equation}
where $p_{ab}^{(i)}$ represents the conditional probability of transitioning from state $a$ to state $b$, calculated via maximum likelihood estimation:
\begin{equation}
p_{ab}^{(k)} = \frac{n_{ab}^{(k)}}{\sum_{m=0}^{1} n_{am}^{(k)}}
\end{equation}
By aggregating the state transitions of all paths within a cluster, the cluster-level probability transition matrix can be obtained, which is listed in Table II.
\section{MODEL IMPLEMENTATION AND VALIDATION}
In this section, we first present the implementation procedure of the channel inference model. Then, we demonstrate its application in map-based channel modeling tasks, where the proposed model leverages canyon information to reconstruct multipath components. Finally, we evaluate the model through statistical characteristic analysis, specifically by comparing the delay spread and AoA spread generated by the proposed model with those extracted from the measurement data.
\subsection{Model Implementation}
According to the proposed model, channel simulations can be implemented by using the parameters listed in Table II. The implementation procedure can be summarized as follows:
%%%%%%%结果图%%%%%%%%%%%%%
\begin{figure*}[t]
    \centering
    \includegraphics[width=\linewidth]{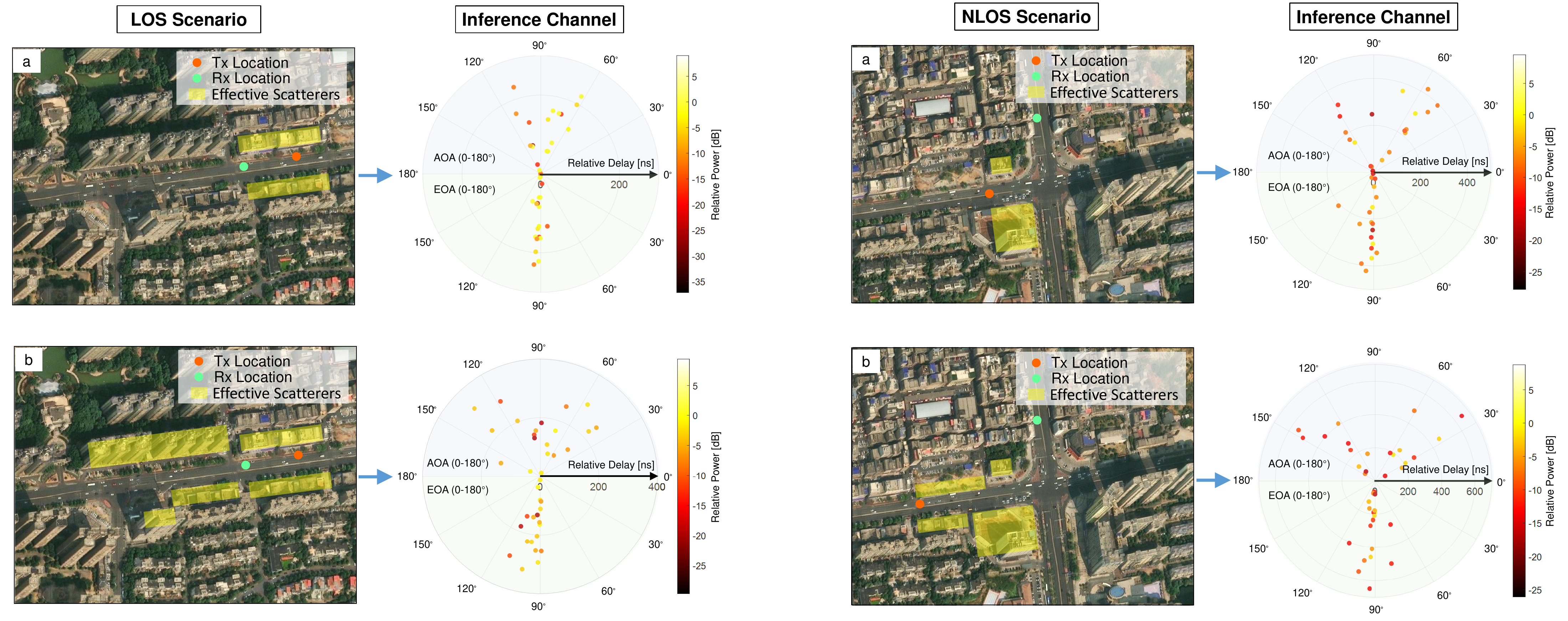}
    \caption{Inference channels at different positions.}
    \label{fig:Result}
\end{figure*}
%%%%%%%%%%%%%%%%%%%%%%%%%%
\begin{itemize}
  \item \textbf{Step 1:} \textit{Set scenario layout.}
  \item \textbf{Step 1a:} \textit{Select the LOS/NLOS state.}
  \item \textbf{Step 1b:} \textit{Determine the predefined parameters}, including the simulation time, antenna array configuration, and environmental parameter $D$. Under LOS conditions, $D$ represents the global set of one-sided canyon widths along the entire route, whereas under NLOS conditions, $D$ represents the global set of one-sided canyon widths only within the NLOS segments.
  \item \textbf{Step 1c:} \textit{Generate the starting positions and velocities of TX, RX.}
  \item \textbf{Step 2:} \textit{Start the simulation loop.}
  \item \textbf{Step 2a:} Calculate the propagation distance $L_\text{LOS}/L_\text{NLOS}$ between TX and RX. For LOS conditions, $L_\text{LOS}$ is the direct distance between TX and RX. For NLOS conditions, first compute the direct distance $L_\text{LOS}$ from TX to the breakpoint, and then compute the distance $L_\text{NLOS}$ from the breakpoint to RX.
  \item \textbf{Step 2b:} Determine the path loss and shadowing using the parameters in Table II. Under LOS conditions, the received power at RX is obtained by inputting $L_\text{LOS}$ into the large-scale parameters of the LOS model. Under NLOS conditions, the received power at RX is determined in two stages: first, input $L_\text{LOS}$ into the LOS large-scale parameters to calculate the signal power at the breakpoint, and then input $L_\text{NLOS}$ into the NLOS large-scale parameters to obtain the received power at RX.
  \item \textbf{Step 2c:} Determine the effective $D(t)$ at time $t$. Here, $D(t)$ denotes the subset of $D$ that influences propagation at time $t$. For LOS conditions, $D(t)$ includes all canyon widths located between TX and RX. For NLOS conditions, $D(t)$ includes all canyon widths between the breakpoint and RX.
  \item \textbf{Step 2d1:} Determine $\psi_X = h_X(D(t);\eta_X)$ using the parameters in Table II.
  \item \textbf{Step 2d2:} Step 2d2: Determine $F_X(\psi_X)$.
  \item \textbf{Step 2d3:} Generate MPCs from the $F_X(\psi_X)$.
  \item \textbf{Step 2e:} Determine birth and death state of MPCs using the parameters in Table II.
  \item \textbf{Step 2f:} \textit{Generate the channel impulse response according to (5).}
  \item \textbf{Step 3:} Update the locations according to the velocity profiles, and then repeat step 2a to step 2f. Note that all MPCs are recomputed after each location has been updated.
  \end{itemize}
Fig. \ref{fig:Model} shows the implementation steps discussed above. It is worth noting that in practical applications, the essential environmental parameter $D$ required for this inference can be easily obtained through simple ranging via aerial photographs or digital maps. To demonstrate how the proposed model is applied, two sets of observation points are designed to showcase the inferred site-specific MPCs, corresponding to the LOS and NLOS scenarios, respectively, as shown in Fig. \ref{fig:Result}. The yellow-lighted buildings in Fig. \ref{fig:Result} denote those that affect the current positions of the transmitter and receiver. In the LOS scenario, as the RX moves farther away from the TX, newly encountered canyons exert an influence on the channel, which manifests as the emergence of additional AoAs, variations in multipath power, and an increase in delay spread. These effects can be clearly observed through the visualization of the site-specific inferred channel.

\subsection{Experimental Setup and Baseline Models}
%%%%%%%CDF验证%%%%%%%%%%%%%
\begin{figure*}[t]
    \centering
    \includegraphics[width=\linewidth]{CDF.pdf}
    \caption{Comparison of CDFs among different models evaluated on the independent validation routes.}
    \label{fig:CDF}
\end{figure*}
%%%%%%%%%%%%%%%%%%%%%%%%%%
To evaluate the performance of the proposed site-specific channel inference model, three representative experimental scenarios were selected for comparative analysis: a LOS scenario, a NLOS scenario, and a mixed scenario encompassing both (LOS \& NLOS). During the model construction phase, the proposed model, the ablation baseline, and the statistical Averaged CDL model uniformly utilized the empirical data collected from the modeling routes illustrated in Fig. \ref{fig:measurement} for parameter extraction and fitting. Subsequently, in the validation phase, the performance of all models was evaluated on the three completely independent validation routes depicted in Fig. \ref{fig:measurement}.

In this experiment, the ablation baseline is designed to verify whether the dynamic physical mapping of the canyon width $D$ yields substantial performance improvements. Specifically, when formulating the mapping function $h_X(D) = \alpha_X D + \beta_X^{(0)}$, this baseline model sets all geometry-dependent slope terms $\alpha_X$ to zero, retaining only the fundamental intercept terms $\beta_X^{(0)}$ extracted from the modeling route data. Under this configuration, the ablation baseline degenerates into a distance-agnostic Model. It forcibly strips away the dynamic evolutionary trend of the channel parameters as the street width expands, outputting solely an initial reference state that is theoretically equivalent to a zero street setback.

The Averaged CDL represents the conventional modeling methodology. The construction of this model strictly adheres to the CDL modeling procedures standardized by 3GPP TR 38.901 \cite{zhu20213gpp}. Initially, the SAGE algorithm is employed to estimate the raw MPCs from the modeling route data. Subsequently, through clustering analysis, these MPCs are grouped into a fixed number of clusters, followed by the calculation of the average relative power, relative delay, and angular spread for each cluster. Ultimately, this model generates a static CDL parameter table, which serves to represent the global average channel state of the region.

Furthermore, the 3GPP TR 38.901 standard model is positioned as the reference baseline in this experiment. Given that the 3GPP model is a generalized statistical model derived from weighting diverse global scenarios, its primary function is to delineate the reference range of channel parameters within a standardized framework. This serves to validate whether the values generated by other models are physically plausible in terms of their order of magnitude rather than to engage in a direct competition for predictive superiority.
\subsection{Performance Evaluation and Results Analysis}
We use two second-order statistics, i.e., the root-mean-square (RMS) delay spread and the angular spreads of AoA, and RMSE path loss to validate our proposed model. We compute the RMS delay spread as the second central moment of the PDP \cite{kanhere2024calibration} and the angular spread according to the definition of Fleury \cite{fleury2002first}. Fig. \ref{fig:CDF} compares the cumulative distribution functions (CDFs) of the parameters between the inference and measured channels. In addition, we use the Kolmogorov-Smirnov-test (KS-test) to measure the distance ($D_{\text{ks}}$) between the distributions of the RMS delay spread and the angular spread of the synthetic data and the measurement data. A smaller $D_{\text{ks}}$ indicates more similarities between the two distributions. As illustrated by the CDF curves in Fig. \ref{fig:CDF} and the statistical results in Table III, the proposed model achieves optimal predictive performance across all parameters, thereby substantiating the effectiveness of explicitly modeling the canyon width $D$.

When comparing the baseline models, two distinctly different performance trends are observed. First, regarding the prediction of the RMS delay spread, the ablation baseline generally outperforms the Averaged CDL model. This is primarily because the delay spread is dictated by the absolute physical distance of signal propagation and is profoundly influenced by the inherent characteristics of the specific environment. Although the ablation baseline forcibly removes the dynamic influence of the street width, it retains the empirical intercept term ($\beta_X^{(0)}$) extracted from the measured data, which is physically equivalent to preserving the initial delay distribution state of that specific environment. In contrast, the Averaged CDL model forcibly averages the delay parameters across various scenarios, inherently destroying the fine-grained characteristics of multipath delays within specific urban canyons.

Second, concerning the RMS AoA spread, the Averaged CDL model consistently outperforms the ablation baseline. This discrepancy arises because AoA characteristics are extremely sensitive to the relative positions of scatterers and heavily rely on the parallax effect between the transceivers and the surrounding buildings. Through the SAGE algorithm and clustering analysis, the Averaged CDL model effectively extracts and preserves the average spatial angular distribution framework of the multipath signals in the region. However, when inferring angles, the ablation baseline completely severs the correlation with the street width $D$. Physically, this equates to completely disregarding the geometric constraints imposed by building locations on the signal arrival angles. Consequently, the model's angular prediction capability degrades drastically, ultimately yielding worse performance than the Averaged CDL model, which at least relies on an averaged angular framework. Finally, regarding the path loss, the proposed model yields RMSE values ranging from 4.44 dB to 6.21 dB across the test scenarios. Since the large-scale fading parameters in the proposed model are fitted using the conventional log-distance method, the path loss RMSE is not intended to demonstrate predictive superiority over baseline models. Instead, these acceptable error margins serve to validate the physical reasonableness of the simulated channel power.

\begin{table*}[htbp]
\centering
\caption{Performance Comparison of Different Methods In Tests}
\label{tab:performance_comparison}
\begin{tabular}{llccc}
\toprule
\textbf{Test Scenario} & \textbf{Method} & \textbf{Delay Spread K-S} & \textbf{AoA Spread K-S} & \textbf{RMSE Path Loss [dB]} \\
\midrule
\multirow{4}{*}{LOS} & Proposed Model & \textbf{0.12} & \textbf{0.12} & \textbf{6.21} \\
                     & 3GPP Model & 0.84 & 1.00 & - \\
                     & Ablation Baseline & 0.22 & 0.57 & - \\
                     & Averaged CDL & 0.82 & 0.26 & - \\
\midrule
\multirow{4}{*}{NLOS} & Proposed Model & \textbf{0.17} & \textbf{0.05} & \textbf{4.44} \\
                      & 3GPP Model & 0.38 & 1.00 & - \\
                      & Ablation Baseline & 0.59 & 0.80 & - \\
                      & Averaged CDL & 0.79 & 0.71 & - \\
\midrule
\multirow{4}{*}{LOS \& NLOS} & Proposed Model & \textbf{0.23} & \textbf{0.19} & \textbf{5.04} \\
                             & 3GPP Model & 0.69 & 0.98 & - \\
                             & Ablation Baseline & 0.81 & 0.62 & - \\
                             & Averaged CDL & 0.76 & 0.29 & - \\
\bottomrule
\end{tabular}
\end{table*}

\section{Conclusion}
In this paper, we propose a site-specific channel inference model based on environmental parameters. The model is developed using extensive measurements conducted in various urban canyon environments. MPCs are extracted from measurement data collected along different streets using a high-resolution algorithm and clustered according to the proposed geometric propagation rules influenced by canyon width, thereby establishing a mapping between the physical environment and statistical characteristics of MPCs. We provide a complete parameterization of the site-specific inference model along with detailed implementation steps. Finally, the model is validated based on measurements. The results show that, in terms of second-order parameters, i.e., RMS delay spread and angle spread, the simulated results exhibit fairly high agreement with measurements, thereby validating scalability and generalization capability of the proposed model.

\bibliographystyle{IEEEtran}
\bibliography{text}

\ifCLASSOPTIONcaptionsoff
  \newpage
\fi

 \end{document}